\documentclass[prl,preprint,showpacs,showkeys,floatfix]{revtex4}
\usepackage{latexsym,amssymb,amsmath,amsfonts}

\begin{document}

\title{Black Hole Thermodynamics and the Factor of 2 Problem}
\author{Terry~Pilling}
\email{Terry.Pilling@ndsu.edu}
\affiliation{Department of Physics,
North Dakota State University,
Fargo, North Dakota 58105}


\begin{abstract}
We show that the recent tunneling formulas 
for black hole radiation in static, spherically symmetric spacetimes 
follow as a consequence of the first law of black hole thermodynamics
and the area-entropy relation based on the radiation temperature. 
A tunneling formula results even if the radiation 
temperature is different from the one originally derived by Hawking 
and this is discussed in the context of the recent factor of 2 problem.
In particular, it is shown that if the radiation temperature is higher 
than the Hawking temperature by a factor of two, thermodynamics 
then leads to a tunneling formula which is exactly the one recently 
found to be canonically invariant.

\end{abstract}

\maketitle

\section{Introduction}

In 1975, Hawking discovered \cite{hawking1975} the remarkable fact 
that black holes radiate a thermal spectrum of particles and that the 
temperature of this
radiation depends on the surface gravity $\kappa$ of the black hole.
This discovery quantified the connections that had been found 
\cite{thermodynamics} between black holes and thermodynamics 
which became the laws of black hole thermodynamics 
\footnote{See section V of Wald (1979) \cite{thermodynamics}.},
summarized as follows:
\begin{description}
\item[(0)]{The temperature of the black hole will be in equilibrium
with the outside temperature.
Since the first law of thermodynamics stated below
gives $T^{-1} = \partial S / \partial E$ as the black hole temperature, 
this must be also equal to the temperature of the radiation. 
}
\item[(1)]{For a static black hole \footnote{This formula has the 
usual generalization for the cases of charged and spinning black 
holes \cite{hawking1976}.} 
of mass $M$, $d M = \frac{\kappa}{8 \pi} dA$, which is analogous to the 
usual first law of thermodynamics $d E = T dS$ if we identify
$\frac{\kappa}{8 \pi} dA = T dS$.
}
\item[(2)]{The entropy of a black hole is proportional to its horizon
area and the sum of the entropy of the black hole and the ordinary
entropy of the matter outside the black hole will never decrease.}
\item[(3)]{The temperature of the black hole is always greater than
or equal to zero.}
\end{description}
The rate of particle emission from the horizon is then proportional
to the change in the black hole entropy 
$\Gamma \sim e^{\beta \Delta E} = e^{\Delta S}$ where $\beta = T^{-1}$.

Since these discoveries it remained somewhat mysterious 
where the particles constituting the radiation come from. 
A physical picture suggested by Hartle and Hawking \cite{hartle1976} 
is that it comes from vacuum fluctuations tunneling through the 
horizon of the black hole and this viewpoint has been adopted by many 
authors. However, the original derivation was not directly connected 
with this viewpoint.
Recently Parikh and Wilczek \cite{parikh2000} and Volovik \cite{volovik1999}
have made the connection by calculating the the 
particle flux in a tunneling picture and showing that the temperatures 
agree with Hawking's original results. 
Since then many calculations \cite{temperatures} have been performed 
using this method verifying that gives the correct temperatures in 
many different backgrounds.

However, Hawking's result that radiation must be emitted by a black 
hole implies a connection with thermodynamics regardless of the actual 
value of the radiation temperature.
So in what follows we will assume the validity of black hole thermodynamics
and show that it leads to a tunneling formula for the radiation but that the 
specific tunneling formula arrived at depends on the definition of 
the entropy of a black hole and so depends on exactly how the 
temperature is related to the surface gravity at the horizon.
In particular, if the radiation temperature were different from that
found by Hawking, the entropy-area law would change accordingly and 
this would change the resulting tunneling formula. 

\section{Tunneling Methods}

There has recently appeared two different tunneling methods
to calculate the Hawking temperature. 
Both formulas come from a semi-classical approximation with a scalar 
field on a curved background to calculate the tunneling amplitude but 
they differ by a factor of 2 in the resulting temperature.
In the first method, called the null geodesic method, one uses Hamilton's 
equation on null geodesics \cite{parikh2000} to calculate the imaginary part 
of the action for particles tunneling across the horizon. This method
gives the same temperatures as Hawking's original calculation. 
The second method which we will call canonically invariant tunneling, one 
uses a particular anzatz for the action and then solves the Hamilton-Jacobi 
equations to find the imaginary part \cite{akhmedov2006,srinivasan1999}. 
This method leads to a formula that is slightly different from the 
null geodesic one in that it is canonically invariant and that it
gives a temperature which is higher than the Hawking temperature by 
a factor of 2 \footnote{Temperatures which differ from the Hawking temperature
have been seen to occur also in fluid analogues to black holes sometimes 
called sonic black holes or dumb holes where the co-tunneling of 
quasiparticles leads to temperatures which differ from the Hawking 
temperature by a factor of $\frac{1}{2}$ \cite{volovik2006}.}.

The idea behind the tunneling methods can be seen by considering the 
space inside, but nearby, the event horizon of the black hole. 
Vacuum fluctuations can then occur and the quantity
$\xi \cdot p + \xi \cdot \overline{p} = 0$ for particle four-momenta
$p$ and $\overline{p}$ and Killing vector $\xi$ must be conserved.
Since $\xi$ is time-like outside the horizon and space-like inside, 
the energy of a particle inside the horizon {\it can be negative} 
\cite{kraus1995}.
The horizon can then move inward with the initial and final horizon
positions constituting the tunneling barrier. 

The tunneling rate for particles through the event horizon using 
the null geodesic method is
\begin{equation}
\label{tunnelingformula}
\Gamma =  e^{-2 \; \text{Im} \int_{r_{in}}^{r_{out}} p_r dr},
\end{equation}
where $p_r$ is the momentum conjugate to $r$ and $r_{in} > r_{out}$ are the
initial and final event horizon radii respectively.
The canonically invariant tunneling method give a rate of
\begin{equation}
\label{canonicalrate}
\Gamma =  e^{- \text{Im} \oint p_r dr}.
\end{equation}
Given a metric with horizon and using Hamilton's equation of motion, 
$\dot{r} = \frac{dE}{dP_r} |_r$, one can calculate $\Gamma$ and
compare with $\Gamma =  e^{- \beta E}$ to extract the radiation temperature
$T = \beta^{-1}$.

In the next section we will show that the first law of black hole 
thermodynamics, along with the usual definitions of the entropy and 
surface gravity, leads to a tunneling formula to leading order in energy.

\section{Thermodynamics and Tunneling}

In order to allow for an arbitrary temperature we will assume
that $T = \eta T_H$ where $T_H$ is the original Hawking temperature
and thus $\eta$ is an arbitrary positive factor. The restriction to
$\eta = 1$ gives the original Hawking temperature, $\eta = 2$ gives
a temperature twice as high as found in the canonically invariant
technique, and $\eta = \frac{1}{2}$ would be the choice for the
fluid analogues with co-tunneling \cite{volovik2006}. 

We consider a general class of static, spherically symmetric spacetimes 
of the form
\begin{equation}
\label{generalmetrics}
ds^2 = - f(r) dt^2 + \frac{dr^2}{g(r)} + r^2 d\Omega^2,
\end{equation}
where the horizon $r = r_H$ is given by $f(r_H) = g(r_H) = 0$.

The metric has a coordinate singularity at the horizon which will remove 
by transforming to Painlev{\' e} coordinates.
let $dt \rightarrow dt - \Delta(r) dr$ and fix $\Delta(r)$ so as to eliminate 
the singularity at $r = r_H$ while requiring the spatial sections at any 
particular time to look like those of Minkowski space (so that the coefficient 
to $dr^2$ is 1), the result is 
\begin{equation}
ds^2 = - f(r) dt^2 + 2 f(r) \sqrt{\frac{1-g(r)}{f(r) g(r)}} dt dr + dr^2 + r^2 d\Omega^2.
\end{equation}
Test particles moving radially in this background will travel along null
geodesics given by
\begin{equation}
\label{nullgeodesics0}
\frac{dr}{dt} \equiv \dot{r} = \sqrt{\frac{f(r)}{g(r)}}\left( \pm 1 - \sqrt{1-g(r)} \right)
\end{equation}
where the positive (negative) sign gives outgoing (incoming) radial geodesics.

The surface gravity of the black hole is defined in terms of a time-like
Killing vector $\mathbf{\xi}$ as 
$\nabla_{\mathbf{\xi}} \mathbf{\xi} = \kappa \; \mathbf{\xi}$,
which reduces to a Christoffel component for our choice of metric
\begin{equation}
\label{surfacegravity}
\kappa = \Gamma^0_{\; 0 0} = \frac{1}{2} \sqrt{\frac{1-g(r)}{f(r) g(r)}} g(r) \frac{d f(r)}{d r} |_{r = r_H}.
\end{equation}

The temperature of a black hole is related to the surface gravity 
$\kappa$ according to $T = \frac{\eta \kappa}{2 \pi}$ where we
have included our factor of $\eta$ described above. 
The entropy \cite{hawking1976,gibbons1977} is then proportional to 
the area, $A$, of the event horizon via the first law of black hole 
thermodynamics as $S = \frac{A}{4 \eta} = \frac{\pi r_H^2}{\eta}$.
The factor of $\eta$ cancels out in the combination $T dS$ to 
reduce to the correct form of the first law of thermodynamics
in terms of surface gravity.
The temperature is then related to the entropy via 
$\frac{ \partial S }{ \partial E} =  \frac{1}{T}$,
where $E$ is the total energy. 

We now take the thermodynamic quantities given above and apply them to 
the region near the horizon.
The derivative of the entropy with respect to energy is
\begin{equation}
\label{hawkingtemp3}
\frac{d S}{d E} = 2 \pi r_H \frac{d r_H}{d E}
\end{equation}
and if the energy of the black hole changes from $E_i$ to $E_f$, the
corresponding change in the entropy is
\begin{equation}
\Delta S = \int_{E_i}^{E_f} 2 \pi r_H \frac{d r_H}{d E} d E.
\end{equation}

The quantities $f(r)$ and $g(r)$ of our metric are zero at the horizon and so 
their expansion in powers of $r-r_H$ are
\begin{equation}
\begin{split}
f(r) &= f'(r_H) (r - r_H) + \cdots, \\
g(r) &= g'(r_H) (r - r_H) + \cdots.
\end{split}
\end{equation}
We will only consider metrics where $f'(r_H)$ and $g'(r_H)$ are non-zero. 
This encompasses most of the important black hole metrics. In the case of 
metrics for which one of $f'(r_H)$ or $g'(r_H)$ is still zero, for example 
extremal black holes \cite{kerner2006}, the following analysis would need 
to be slightly modified.
By inserting these expressions into (\ref{nullgeodesics0}) and choosing
the positive sign since we are looking at the `outward' geodesics 
we can write the near horizon radial geodesic equation as
\begin{equation}
\label{nullgeodesic}
\dot{r} = \frac{1}{2} \sqrt{f'(r_H)g'(r_H)} (r - r_H).
\end{equation}
Since $f'(r_H)$ and $g'(r_H)$ are non-zero we can invert this equation to give
\begin{equation}
\label{deltafunction}
r - r_H = \frac{2 \dot{r}}{\sqrt{f'(r_H)g'(r_H)}}.
\end{equation}

The surface gravity is found to be 
$\kappa = \frac{1}{2} \sqrt{f'(r_H) g'(r_H)}$,
which gives the temperature as 
\begin{equation}
\label{hawkingtemp2}
T = \frac{\eta \sqrt{f'(r_H) g'(r_H)}}{4 \pi}.
\end{equation}
For a small path from $r_i$ to $r_f$ containing $r_H$ we make
the connection to tunneling with the identity
\begin{equation}
\text{Im} \int_{r_i}^{r_f} \frac{1}{r-r_H} dr = - \pi 
\end{equation}
so that the horizon radius can be written as
\begin{equation}
r_H = - \text{Im} \frac{1}{\pi} \int_{r_i}^{r_f} \frac{r_H}{r-r_H} dr 
\end{equation}
and thus we can write our entropy change in the mathematically equivalent 
form
\begin{equation}
\Delta S = - \text{Im} \int_{E_i}^{E_f} \int_{r_i}^{r_f} \frac{2 r_H}{r-r_H} \frac{d r_H}{d E} dr dE.
\end{equation}
Now using (\ref{deltafunction}) this becomes
\begin{equation}
\label{deltaS}
\Delta S = - \text{Im} \int_{E_i}^{E_f} \int_{r_i}^{r_f} \sqrt{f'(r_H) g'(r_H)} \frac{r_H}{\dot{r}} \frac{d r_H}{d E} dr dE.
\end{equation}
Equations (\ref{hawkingtemp3}), and (\ref{hawkingtemp2}) along with the first
law give
\begin{equation}
r_H \frac{d r_H}{d E} = \frac{2}{\eta \sqrt{f'(r_H) g'(r_H)}}
\end{equation}
and our change in entropy (\ref{deltaS}) becomes
\begin{equation}
\Delta S = - \frac{2}{\eta} \; \text{Im} \int_{m}^{m-\omega} \int_{r_i}^{r_f} \frac{dr}{\dot{r}} dE.
\end{equation}
where we have written the initial energy as the mass $m$ and the final
energy as $m - \omega$ where $\omega$ is interpreted as the energy radiated.
To first order in $\omega$ the right hand side is identical, aside from our 
factor of $\eta$, to the 
expression for a particle tunneling through the event horizon on a null 
geodesic in the $s$-wave WKB approximation given 
in (\ref{tunnelingformula}) above. Therefore we can interpret
$\omega$ as the energy of a tunneling particle. 
On the other hand, from the tunneling point of view 
\begin{equation}
\dot{r} = \dot{r}(r,m-\omega) = \dot{r}(r, m) + {\cal O}(\omega)
\end{equation}
where the higher order corrections lead to non-thermal corrections 
to the black body Hawking spectrum.
Thus we have seen that, to first order in the energy of the 
tunneling particle, the laws of black hole thermodynamics lead
directly to the following relation
\begin{equation}
\Gamma \sim e^{\Delta S} = e^{- \frac{2}{\eta} \; \text{Im} \int_{r_{in}}^{r_{out}} p_r dr},
\end{equation}
in which $\eta = 1$ gives the null geodesic tunneling formula
\cite{parikh2000,volovik1999}
and $\eta = 2$ is equivalent to the canonically invariant tunneling formula 
$\Gamma \sim e^{- \text{Im } \oint p dr}$ \cite{akhmedov2006,chowdhury2007}.
It is important to notice that the formula follows mathematically
from the entropy-area relation without using quantum field theory. 
The semi-classical field theory derivations from
tunneling give similar formulas which differ in the choice 
of $\eta$. 
Once a tunneling formula has been derived, the temperature can be read 
off of it through $\eta$ and so the tunneling formulas amount to predictions 
for the factor of $\eta$. 

As it stands we have two different tunneling formulas implying two
different temperatures. In order to see which formula is correct, 
one needs to calculate the entropy of a black hole in an independent 
way and then compare with $S = \frac{A}{4 \eta}$ to extract $\eta$. 
The canonically invariant tunneling prediction is that $\eta = 2$ whereas
the null geodesic method predicts $\eta = 1$.

\section{Some Consequences of the Tunneling Formalism}

We have found that the integrand found in the tunneling method is always 
related to the entropy as 
$\int \frac{\partial S}{\partial m} dm$ \cite{fang2006}. 
This relation is not obvious from the tunneling picture alone, but with the 
connection to thermodynamics established the reason is just the one 
given by Hawking when he wrote down the same formula 
in 1976 \cite{hawking1976}.

Given our present demonstration one can now calculate the self-gravitation 
corrections to the Hawking temperature in an easy way, directly from the 
expression for the entropy. 
For example, in the Schwarzschild case the entropy is 
$S = \frac{\pi r_H^2}{\eta}$,
where $r_H = 2m$. Thus, writing $m' = m - \omega$ we have
\begin{equation}
\Delta S 
= \int \frac{\partial S}{\partial m'} (-d\omega)
= - \frac{8 \pi \omega}{\eta} \left( m - \frac{\omega}{2} \right)
\end{equation}
which, for $\eta = 1$, is the expression derived in \cite{parikh2000}.
As a less trivial example the Reissner-Nordstr{\" o}m black 
hole with line element
\begin{equation}
ds^2 = -  f(r) \; dt^2 + f(r)^{-1} dr^2 + r^2 d\Omega^2
\end{equation}
where $f(r) = \left(1-\frac{2m}{r} + \frac{q^2}{r^2} \right)$ and the 
radii of the inner and outer horizons are given by 
$r_\pm = m \pm \sqrt{m^2 - q^2}$.
The entropy is $S = \frac{\pi r_+^2}{\eta}$ and so with $m' = m - \omega$ 
we find a formula for the change in entropy including self-gravitation 
corrections 
\begin{equation}
\begin{split}
\Delta S 
&= - \frac{2 \pi}{\eta} \Bigl\{ m \sqrt{m^2-q^2} + 2\omega(m-\omega/2) \\
&\hspace{2cm} - (m-\omega) \sqrt{(m-\omega)^2 - q^2} \Bigr\}.
\end{split}
\end{equation}
From which the rate $\Gamma \sim e^{\Delta S}$ follows
giving the radiation temperature for $\omega = 0$ as
\begin{equation}
T = \frac{\eta}{2 \pi} \frac{\sqrt{m^2 - q^2}}{\left(m + \sqrt{m^2-q^2}\right)^2}
\end{equation}

Notice that in the extreme case $q^2 = m^2$ the Hawking temperature vanishes,
which implies that the tunneling amplitude also vanishes. This is consistent
with cosmic censorship. However we should check that this is really true, since
our derivation is not valid in the extremal case due to the fact that the
integrand is no longer just a simple pole. 
Another reason to check the derivation for the extremal case is that, when 
self gravity corrections are included in the extremal limit our naive formula 
for the amplitude no longer seems to vanish. 
This may incline one to believe that cosmic censorship may be violated by 
higher order quantum corrections.
In fact one can easily see that it does still vanish. In the near extremal 
case, with outer horizon at $r = r_H$ and inner horizon at $r = r_H - \epsilon$
the tunneling amplitude is $\Gamma \sim e^{-\alpha/\epsilon}$ for constant
$\alpha > 0$ and thus vanishes in the extremal limit. In the exact extremal
case one can use the fact that
\begin{equation}
\frac{d f(z_0)}{d z_0} = \oint_C \frac{f(z)}{(z-z_0)^2} dz
\end{equation}
to see that the integral vanishes exactly. This occurs before we even
complete the $dE$ integral and so quantum corrections should not alter the
result and we find that cosmic censorship is not violated by tunneling.

\section{The Factor of 2 Problem}

We have shown that to first order in the energy of the tunneling particle 
the tunneling picture follows from the first law of black hole 
thermodynamics and the entropy-area relation. The specific form 
of the tunneling formula then follows from the specific value of the 
radiation temperature. The choice $\eta = 1$ gives the null
geodesic tunneling formula, used by many authors, corresponding to 
the original Hawking temperature.

However, Chowdhury \cite{chowdhury2007} in a recent analysis of the
null geodesic tunneling formula has shown that it is {\it not invariant under 
canonical transformations}, but that the same formula with a factor of 
1/2 in the exponent {\it is canonically invariant}. It seems clear physically,
as Chowdhury argues, that one should use the canonically invariant formula
$\Gamma \sim e^{- \text{Im } \oint p dr}$. This formula reduces 
to our $\eta = 2$ formula for black hole horizons because a 
horizon is a one-way tunneling barrier and so the transmission
coefficient for the half of the path which is directed inward is
equal to 1 \footnote{See equation (23) in \cite{chowdhury2007}.}.
This is interesting in light of the recent results of
Akhmedov {\it et.al.} who showed that the temperatures found using 
the canonically invariant tunneling formula differ by a factor of 2 from 
the Hawking results in several backgrounds \cite{akhmedov2006}. 
Nakamura \cite{nakamura2007} has recently offered an explanation of this 
difference in terms of the alternate vacuum used by the tunneling methods
and this may solve the factor of 2 problem, but if so it it raises a further 
question. 
How is it that, when the null geodesic method (being a tunneling method) 
uses the alternate vacuum, it still reproduces the original Hawking 
temperatures without the factor of 2? 

Our present demonstration shows that the paradox can be resolved
if the black hole temperature really is a factor of two higher 
than that originally given by Hawking.
This would mean $\eta = 2$ and the black hole entropy would be 
$S = A/8$ instead of the usual relation, leading
to a different formula for the tunneling. In fact,
the formula that results is exactly the canonically
invariant formula, $\Gamma \sim e^{- \text{Im } \oint p dr}$, given by 
Akhmedov {\it et.al} and Chowdhury. 

We conclude that setting $\eta = 2$ makes the tunneling formula 
consistent with Nakamura's result for tunneling vacua, consistent with 
Chowdhury's result on the canonical invariance and consistent with the 
tunneling derivation of Akhmedov {\it et.al.}
However, the resulting formula contradicts Hawking's 
original value of the black hole radiation temperature.
The tunneling methods each give a different prediction for the
factor of $\eta$ in the entropy-area relation and therefore
it is necessary to calculate the black hole entropy in a 
completely independent way to find out which is correct.
There are different groups working on calculating the 
entropy using stringy or loopy microstates \cite{ashtekar1998,akhmedov2000} 
and it is hoped that they will independently fix $\eta$.

\section{Acknowledgments}

I would like to thank Emil Akhmedov, Douglas Singleton 
and Grigori Volovik for helpful comments.

\end{document}